\documentstyle[preprint,epsfig,tighten,floats,aps,amsfonts,amssymb]{revtex}
\def\C{{\cal C}}
\def\D{{\cal D}}
\def\E{{\cal E}}
\def\G{{\cal G}}
\def\H{{\cal H}}
\def\Hih{{\cal H}_{\rm HI}}

\def\S{{\Sigma}}

\def\SU{{\rm SU}}
\def\U{{\rm U}}
\def\SL{{\rm SL}}
\def\lp{\ell_{\rm Pl}}

\def\be{\begin{equation}}
\def\ee{\end{equation}}
\def\ba{\begin{eqnarray}}
\def\ea{\end{eqnarray}}

\begin{document}

\title{Quantum Geometry and Gravity: Recent Advances}
\author{Abhay\ Ashtekar} 
\address{Center for Gravitational Physics and Geometry, \\
Physics Department, Penn State, University Park, PA 16802, USA}

\maketitle
\begin{abstract}
Over the last three years, a number of fundamental physical issues
were addressed in loop quantum gravity. These include: A
statistical mechanical derivation of the horizon entropy,
encompassing astrophysically interesting black holes as well as
cosmological horizons; a natural resolution of the big-bang
singularity; the development of spin-foam models which provide
background independent path integral formulations of quantum
gravity and `finiteness proofs' of some of these models; and, the
introduction of semi-classical techniques to make contact between
the background independent, non-perturbative theory and the
perturbative, low energy physics in Minkowski space. These
developments spring from a detailed quantum theory of geometry
that was systematically developed in the mid-nineties and have
added a great deal of optimism and intellectual excitement to the
field.

The goal of this article is to communicate these advances in
general physical terms, accessible to researchers in all areas of
gravitational physics represented in this conference.

\end{abstract}

\section{Introduction}
\label{s1}

Let us begin by recalling some of the central conceptual and
physical questions of quantum gravity.

$\bullet$ \textit{Big-Bang and other singularities:} It is widely
believed that the prediction of a singularity, such as the
big-bang of classical general relativity, is primarily a signal
that the theory has been pushed beyond the domain of its validity.
A key question to any quantum gravity theory, then, is: What
replaces the big-bang?  Qualitatively, classical geometry may be a
mean field like `magnetization', which provides an excellent
macroscopic description of a ferromagnet. However, at the Curie
temperature, magnetization goes to zero and susceptibility
diverges. But there is no physical infinity; we simply have to
turn to the correct \textit{microscopic} description in terms of
spin-systems to describe physics. Does something similar happen at
the big-bang and other singularities? Is there a mathematically
consistent description of the quantum state of the universe which
replaces the classical big-bang? What is the analog of the
microscopic spin-system that underlies magnetism? What can we say
about the `initial conditions', i.e., the quantum state of
geometry and matter that correctly describes the big-bang? If they
have to be imposed externally, is there a \textit{physical}
guiding principle?

$\bullet$ \textit{Black holes:} In the early seventies, using
imaginative thought experiments, Bekenstein argued that black
holes must carry an entropy proportional to their area. About the
same time, Bardeen, Carter and Hawking (BCH) showed that black
holes in equilibrium obey two basic laws, which have the same form
as the zeroth and the first laws of thermodynamics, provided one
equates the black hole surface gravity $\kappa$ to some multiple
of the temperature $T$ in thermodynamics and the horizon area
$a_{\rm hor}$ to a corresponding multiple of the entropy $S$.
However, at first this similarity was thought to be only a formal
analogy because the BCH analysis was based on \textit{classical}
general relativity and simple dimensional considerations show that
the proportionality factors must involve Planck's constant
$\hbar$. Two years later, using quantum field theory on a black
hole background space-time, Hawking showed that black holes in
fact radiate quantum mechanically as though they are black bodies
at temperature $T = \hbar\kappa/2\pi$. Using the analogy with the
first law, one can then conclude that the black hole entropy
should be given by $S_{\rm BH} = a_{\rm hor}/4G\hbar$. This
conclusion is striking and deep because it brings together the
three pillars of fundamental physics ---general relativity,
quantum theory and statistical mechanics. However, the argument
itself is a rather hodge-podge mixture of classical and
semi-classical ideas, reminiscent of the Bohr theory of atom. A
natural question then is: what is the analog of the more
fundamental, Pauli-Schr\"odinger theory of the Hydrogen atom? More
precisely, what is the statistical mechanical origin of black hole
entropy? What is the nature of a quantum black hole and what is
the interplay between the quantum degrees of freedom responsible
for entropy and the exterior curved geometry? Can one derive the
Hawking effect from first principles of quantum gravity? Is there
an imprint of the classical singularity on the final quantum
description, e.g., through `information loss' ?

$\bullet$ \textit{Planck scale physics and the low energy world:}
Perhaps the central lesson of general relativity is that
\textit{gravity is geometry}. There is no longer a background
metric, no inert stage on which dynamics unfolds. Geometry itself
is dynamical. Therefore, one expects that a fully satisfactory
quantum gravity theory would also be free of a background
space-time geometry. However, of necessity, a background
independent description must use physical concepts and
mathematical tools that are quite different from those of the
familiar, low energy physics. A major challenge then is to show
that this low energy description does arise from the pristine,
Planckian world in an appropriate sense. In this `top-down'
approach, does the fundamental theory admit a ``sufficient
number'' of semi-classical states? Do these semi-classical sectors
provide enough of a background geometry to anchor low energy
physics? can one recover the familiar description? Furthermore,
can one pin point why the standard `bottom-up' perturbative
approach fail? That is, what is the essential feature which makes
the fundamental description mathematically coherent but is absent
in the standard perturbative quantum gravity?

Of course, this is by no means a complete list of challenges.
There are many others: the issue of time, of measurement theory
and the associated questions of interpretation of the quantum
framework, the issue of diffeomorphism invariant observables and
practical methods of computing their properties, practical methods
of computing time evolution and S-matrices, exploration of the
role of topology and topology change, \ldots. The purpose of this
report is to summarize recent advances in the non-perturbative
approach based on quantum geometry which has led to illuminating
answers to many of these questions and opened-up avenues to
address others. The plenary session of this conference also
covered simplicial quantum gravity and string theory which, in a
certain sense, complement our approach. And there are other
approaches as well, ranging from twistors and causal sets to
Euclidean path integrals. Unfortunately, due to space limitation,
I will not be able to discuss these; indeed, even within the
approach I focus on, I can discuss only a few illustrative
examples. I apologize in advance to the authors whose very
interesting contributions could not be referred to in this brief
report.

\section{A bird's eye view of loop quantum gravity}
\label{s2}

In this section, I will briefly summarize the salient features and
current status of loop quantum gravity. The emphasis is on
structural and conceptual issues; detailed treatments can be found in
references [1-9] and papers they refer to.

\subsection{Viewpoint}
\label{s2.1}

In this approach, one takes the central lesson of general
relativity seriously: gravity \textit{is} geometry whence, in a
fundamental theory, there should be no background metric. In
quantum gravity, geometry and matter should \textit{both} be `born
quantum mechanically'. Thus, in contrast to approaches developed
by particle physicists, one does not begin with quantum matter on
a background geometry and use perturbation theory to incorporate
quantum effects of gravity. There \textit{is} a manifold but no
metric, or indeed any other physical fields, in the background.%
\footnote{In 2+1 dimensions, although one begins in a completely
analogous fashion, in the final picture one can get rid of the
background manifold as well. Thus, the fundamental theory can be
formulated combinatorially \cite{loop,books}. To achieve this goal
in 3+1 dimensions, one needs a much better understanding of the
theory of (intersecting) knots in 3 dimensions.}

In the classical gravity, Riemannian geometry provides the
appropriate mathematical language to formulate the physical,
kinematical notions as well as the final dynamical equations. This
role is now taken by \textit{quantum} Riemannian geometry,
discussed below. In the classical domain, general relativity
stands out as the best available theory of gravity, some of whose
predictions have been tested to an amazing accuracy, surpassing
even the legendary tests of quantum electrodynamics. Therefore, it
is natural to ask: \textit{Does quantum general relativity,
coupled to suitable matter} (or supergravity, its supersymmetric
generalization) \textit{exist as consistent theories
non-perturbatively?} There is no a priori implication that such a
theory would be the final, complete description of Nature.
Nonetheless, this is a fascinating open question, at least at the
level of mathematical physics.

In the particle physics circles, the answer is often assumed to be
in the negative, not because there is concrete evidence against
non-perturbative quantum gravity, but because of an analogy to the
theory of weak interactions. There, one first had a 4-point
interaction model due to Fermi which works quite well at low
energies but which fails to be renormalizable. Progress occurred
not by looking for non-perturbative formulations of the Fermi
model but by replacing the model by the Glashow-Salam-Weinberg
renormalizable theory of electro-weak interactions, in which the
4-point interaction is replaced by $W^\pm$ and $Z$ propagators.
Therefore, it is often assumed that perturbative
non-renormalizability of quantum general relativity points in a
similar direction. However this argument overlooks the crucial
fact that, in the case of general relativity, there is a
qualitatively new element. Perturbative treatments pre-suppose
that the space-time can be assumed to be a continuum \textit{at
all scales} of interest to physics under consideration. In the
gravitational case, the scale of interest is given by the Planck
length $\lp$ and there is no physical basis to pre-suppose that
the continuum picture should be valid down to that scale. The
failure of the standard perturbative treatments may simply be due
to this grossly incorrect assumption and a non-perturbative
treatment which correctly incorporates the physical
micro-structure of geometry may well be free of these
inconsistencies.

As indicated above, even if quantum general relativity did exist
as a mathematically consistent theory, there is no a priori reason
to assume that it would be the `final' theory of all known
physics. In particular, as is the case with classical general
relativity, while requirements of background independence and
general covariance do restrict the form of interactions between
gravity and matter fields and among matter fields themselves, the
theory would not have a built-in principle which
\textit{determines} these interactions. Put differently, such a
theory would not be a satisfactory candidate for unification of
all known forces. However, just as general relativity has had
powerful implications in spite of this limitation in the classical
domain, quantum general relativity should have qualitatively new
predictions, pushing further the existing frontiers of physics.
Indeed, unification does not appear to be an essential criterion
for usefulness of a theory even in other interactions. QCD, for
example, is a powerful theory even though it does not unify strong
interactions with electro-weak ones. Furthermore, the fact that we
do not yet have a viable candidate for the grand unified theory
does not make QCD any less useful.

\subsection{Quantum Geometry}
\label{s2.2}

Although there is no natural unification of dynamics of all
interactions in loop quantum gravity, it does provide a
kinematical unification. More precisely, in this approach one
begins by formulating general relativity in the mathematical
language of connections, the basic variables of gauge theories of
electro-weak and strong interactions. Thus, now the configuration
variables are not metrics as in Wheeler's geometrodynamics, but
certain spin connections; the emphasis is shifted from distances
and geodesics to holonomies and Wilson loops \cite{books}.
Consequently, the basic kinematical structures are the same as
those used in gauge theories. A key difference, however, is that
while a background space-time metric is available and crucially
used in gauge theories, there are no background fields whatsoever
now. This absence is forced on us by the requirement of
diffeomorphism invariance (or `general covariance' ).

This is a key difference and it causes a host of conceptual as
well as technical difficulties in the passage to quantum theory.
For, most of the techniques used in the familiar, Minkowskian
quantum theories are deeply rooted in the availability of a flat
back-ground metric. It is this structure that enables one to
single out the vacuum state, perform Fourier transforms to
decompose fields canonically in to creation and annihilation
parts, define masses and spins of particles and carry out
regularizations of products of operators. Already when one passes
to quantum field theory in \textit{curved} space-times, extra work
is needed to construct mathematical structures that can adequately
capture underlying physics. In our case, the situation is much
more drastic: there is no background metric what so ever!
Therefore new physical ideas and mathematical tools are now
necessary. Fortunately, they were constructed by a number of
researchers in the mid-nineties and have given rise to a detailed
quantum theory of geometry \cite{qg1,qg2,qg3,bi}.

Because the situation is conceptually so novel and because there
are no direct experiments to guide us, reliable results require a
high degree of mathematical precision to ensure that there are no
hidden infinities.  Achieving this precision has been a priority
in the program. Thus, while one is inevitably motivated by
heuristic, physical ideas and formal manipulations, the final
results are mathematically rigorous. In particular, due care is
taken in constructing function spaces, defining measures and
functional integrals, regularizing products of field operator, and
calculating eigenvectors and eigenvalues of geometric operators.
Consequently, the final results are all free of divergences,
well-defined, and respect the background independence and
diffeomorphism invariance.

Let me now turn to specifics. Our basic configuration variable is
an $\SU(2)$-connection, $A_a^i$ on a 3-manifold $\S$ representing
`space' and, as in gauge theories, the momenta are the `electric
fields' $E^a_i$. However, in the present gravitational context,
they acquire an additional meaning: they can be naturally
interpreted as orthonormal triads (with density weight $1$) and
determine the dynamical, Riemannian geometry of $\S$. Thus, in
contrast to Wheeler's geometrodynamics, the Riemannian structures,
including the positive-definite metric on $\S$, is now built from
\textit{momentum} variables. The basic kinematic objects are
holonomies of $A_a^i$, which dictate how spinors are parallel
transported along curves, and the triads $E^a_i$, which determine
the geometry of $\S$. (Matter couplings to gravity have also been
studied extensively \cite{class,books}.)

In the quantum theory, the fundamental excitations of geometry are
most conveniently expressed in terms of holonomies
\cite{loop,qg1}. They are thus \textit{one-dimensional,
polymer-like} and, in analogy with gauge theories, can be thought
of as `flux lines of the electric field'. More precisely, they
turn out to be flux lines of areas, the simplest gauge invariant
quantities constructed from $E^a_i$: an elementary flux line
deposits a quantum of area on any 2-surface $S$ it intersects.
Thus, if quantum geometry were to be excited along just a few flux
lines, most surfaces would have zero area and the quantum state
would not at all resemble a classical geometry. This state would
be analogous, in Maxwell theory, to a `genuinely quantum
mechanical state' with just a few photons. In the Maxwell case,
one must superpose photons coherently to obtain a semi-classical
state that can be approximated by a classical electromagnetic
field. Similarly, here, semi-classical geometries can result only
if a huge number of these elementary excitations are superposed in
suitable dense configurations \cite{sc,fock}. The state of quantum
geometry around you, for example, must have so many elementary
excitations that $\sim 10^{68}$ of them intersect the sheet of
paper you are reading. Even in such states, the geometry is still
distributional, concentrated on the underlying elementary flux
lines; but if suitably coarse-grained, it can be approximated by a
smooth metric. Thus, the continuum picture is only an
approximation that arises from coarse graining of semi-classical
states.

These quantum states span a specific Hilbert space $\H$ consisting
of wave functions of connections which are square integrable with
respect to a natural, diffeomorphism invariant measure \cite{qg1}.
This space is very large. However, it can be conveniently
decomposed in to a family of orthonormal, \textit{finite}
dimensional sub-spaces $\H = \oplus_{\gamma, \vec{j}}\,\,
\H_{\gamma, \vec{j}}$, labelled by graphs $\gamma$ each edge of
which itself is labelled by a spin (i.e., half-integer) ${j}$
\cite{qg2}. (The vector ${\vec j}$ stands for the collection of
half-integers associated with all edges of $\gamma$.) One can
think of $\gamma$ as a `floating lattice' in $\S$ ---`floating'
because its edges are arbitrary, rather than `rectangular'.
(Indeed, since there is no background metric on $\S$, a
rectangular lattice has no invariant meaning.) Mathematically,
$\H_{\gamma,\vec{j}}$ can be regarded as the Hilbert space of a
spin-system. These spaces are extremely simple to work with; this
is why very explicit calculations are feasible. Elements of
$\H_{\gamma,\vec{j}}$ are referred to as \textit{spin-network
states} \cite{qg2}.

As one would expect from the structure of the classical theory,
the basic quantum operators are the holonomies $\hat{h}_p$ along
paths $p$ in $\S$ and the triads $\hat{E}^a_i$ \cite{qg3}. (Both
sets are densely defined and self-adjoint on $\H$.) Furthermore, a
striking result is that \textit{all eigenvalues of the triad
operators are discrete.} This key property is, in essence, the
origin of the fundamental discreteness of quantum geometry. For,
just as the classical Riemannian geometry of $\S$ is determined by
the triads $E^a_i$, all Riemannian geometry operators  ---such as
the area operator $\hat{A}_S$ associated with a 2-surface $S$ or
the volume operator $\hat{V}_R$ associated with a region $R$---
are constructed from $\hat{E}^a_i$. However, since even the
classical quantities $A_S$ and $V_R$ are non-polynomial
functionals of the triads, the construction of the corresponding
$\hat{A}_S$ and $\hat{V}_R$ is quite subtle and requires a great
deal of care. But their final expressions are rather simple
\cite{qg3}.

In this regularization, the underlying background independence
turns out to be a blessing. For, diffeomorphism invariance
constrains the possible forms of the final expressions
\textit{severely} and the detailed calculations then serve
essentially to fix numerical coefficients and other details. Let
us illustrate this point with the example of the area operators
$\hat{A}_S$. Since they are associated with 2-surfaces $S$ while
the states are 1-dimensional excitations, the diffeomorphism
covariance requires that the action of $\hat{A}_S$ on a state
$\Psi_{\gamma, \vec{j}}$ must be concentrated at the intersections
of $S$ with $\gamma$. The detailed expression bears out this
expectation: the action of $\hat{A}_S$ on $\Psi_{\gamma, \vec{j}}$
is dictated simply by the spin labels $j_I$ attached to those
edges of $\gamma$ which intersect $S$. For all surfaces $S$ and
3-dimensional regions $R$ in $\S$,  $\hat{A}_S$ and $\hat{V}_R$
are densely defined, self-adjoint operators. \textit{All their
eigenvalues are discrete} \cite{qg3}. Naively, one might expect
that the eigenvalues would be uniformly spaced, given by, e.g.,
integral multiples of the Planck area or volume.  This turns out
\textit{not} to be the case; the distribution of eigenvalues is
quite subtle. In particular, the eigenvalues crowd rapidly as
areas and volumes increase. In the case of area operators, the
complete spectrum is known in a \textit{closed form}, and the
first several hundred eigenvalues have been explicitly computed
numerically. For a large eigenvalue $a_n$, the separation $\Delta
a_n = a_{n+1} -a_n$ between consecutive eigenvalues decreases
exponentially: $\Delta a_n \, \le\, \lp^2 \, \exp -(\sqrt{a_n}/\lp
)$! Because of such strong crowding, the continuum approximation
becomes excellent quite rapidly just a few orders of magnitude
above the Planck scale. At the Planck scale, however, there is a
precise and very specific replacement. This is the arena of
quantum geometry. The premise is that the standard perturbation
theory fails because it ignores this fundamental discreteness.

There is however a further subtlety \cite{class,bi}. This
non-perturbative quantization has a one parameter family of
ambiguities labelled by $\gamma > 0$. This $\gamma$ is called the
Barbero-Immirzi parameter and is rather similar to the well-known
$\theta$-parameter of QCD. In QCD, a single classical theory gives
rise to inequivalent sectors of quantum theory, labelled by
$\theta$. Similarly, $\gamma$ is classically irrelevant but
different values of $\gamma$ correspond to unitarily inequivalent
representations of the algebra of geometric operators. The overall
mathematical structure of all these sectors is very similar; the
only difference is that the eigenvalues of all geometric operators
scale with $\gamma$. For example, the simplest eigenvalues of the
area operator $\hat{A}_S$ in the $\gamma$ quantum sector is given
by %
\footnote{In particular, the lowest non-zero eigenvalue of area
operators is proportional to $\gamma$. This fact has led to a
misunderstanding: in circles outside loop quantum gravity,
$\gamma$ is sometimes thought of as a regulator responsible for
discreteness of quantum geometry. As explained above, this is
\textit{not} the case; $\gamma$ is analogous to the QCD $\theta$
and quantum geometry is discrete in \textit{every} permissible
$\gamma$-sector. Note also that, at the classical level, the
theory is equivalent to general relativity only if $\gamma$ is
\textit{positive}; if one sets $\gamma= 0$ by hand, one can not
recover even the kinematics of general relativity. Similarly, at
the quantum level, setting $\gamma=0$ would lead to a meaningless
theory in which \textit{all} eigenvalues of geometric operators
vanish identically.}
\be \label{2.1} a_{\{j\}} = 8\pi\gamma \lp^2 \, \sum_I \sqrt{j_I
(j_I +1)} \ee
where $\{j\}$ is a collection of 1/2-integers $j_I$, with $I =
1,\ldots N$ for some $N$. Since the representations are unitarily
inequivalent, as usual, one must rely on Nature to resolve this
ambiguity: Just as Nature must select a specific value of $\theta$
in QCD, it must select a specific value of $\gamma$ in loop
quantum gravity. With one judicious experiment
---e.g., measurement of the lowest eigenvalue of the area operator
$\hat{A}_S$ for a 2-surface $S$ of any given topology--- we could
determine the value of $\gamma$ and fix the theory. Unfortunately,
such experiments are hard to perform! However, we will see in
Section \ref{s3.2} that the Bekenstein-Hawking formula of black
hole entropy provides an indirect measurement of this lowest
eigenvalue of area for the 2-sphere topology and can therefore be
used to fix the value of $\gamma$.

\subsection{Quantum dynamics}
\label{s2.3}

Quantum geometry provides a mathematical arena to formulate
non-perturbative dynamics of candidate quantum theories of
gravity, without any reference to a background classical geometry.
In the case of general relativity, it provides the tools to write
down quantum Einstein's equations in the Hamiltonian approach and
calculate transition amplitudes in the path integral approach.
Until recently, effort was focussed primarily on the Hamiltonian
approach. However, over the last three years, path integrals
---called \textit{spin foams}--- have drawn a great deal of
attention. This work has led to fascinating results suggesting
that, thanks to the fundamental discreteness of quantum geometry,
path integrals defining quantum general relativity may be finite.
These developments will be discussed in some detail in Section
\ref{s4.1}. In this Section, I will summarize the status of the
Hamiltonian approach. For brevity, I will focus on source-free
general relativity, although there has been considerable work also
on matter couplings.

For simplicity, let me suppose that the `spatial' 3-manifold $\S$
is compact. Then, in any theory without background fields,
Hamiltonian dynamics is governed by constraints. Roughly, this is
because, in these theories, diffeomorphisms correspond to gauge in
the sense of Dirac. Recall that, on the Maxwell phase space, gauge
transformations are generated by the functional $\D_a E^a$ which
is constrained to vanish on physical states due to Gauss law.
Similarly, on phase spaces of background independent theories,
diffeomorphisms are generated by Hamiltonians which are
constrained to vanish on physical states. In the case of general
relativity, there are three sets of constraints. The first set
consists of the three Gauss equations
$${\G}_i:= \D_a \, E^a_i =0, $$
which, as in Yang-Mills theories, generates internal $\SU(2)$
rotations on the connection and the triad fields. The second set
consists of a co-vector (or diffeomorphism) constraint
$${\C}_b :=E^a F_{ab} = 0, $$
which generates spatial diffeomorphism on $\S$ (modulo internal
rotations generated by ${\G}_i$). Finally, there is the key scalar
(or Hamiltonian) constraint
$${\cal S} := \epsilon^{ijk} E^a_i E^b_j F_{ab}{k} + \ldots = 0$$
which generates time-evolutions. (The $\ldots$ are extrinsic
curvature terms, expressible as Poisson brackets of the
connection, the total volume constructed from triads and the first
term in the expression of ${\cal S}$ given above. We will not need
their explicit forms.) Our task in quantum theory is three-folds:
i) Elevate these constraints (or their `exponentiated versions')
to well-defined operators on the kinematic Hilbert space $\H$; ii)
Select physical states by asking that they be annihilated by these
constraints; iii) introduce an inner-product and interesting
observables, and develop approximation schemes, truncations, etc
to explore physical consequences.

Step i) has been completed. Since the action of the Gauss and the
co-vector constraints have a simple geometrical meaning,
completion of i) in these cases is fairly straightforward. For the
scalar constraint, on the other hand, there are no such guiding
principles whence the procedure is subtle. In particular, specific
regularization choices have to be made. Consequently, the answer
is not unique. At the present stage of the program, such
ambiguities are inevitable; one has to consider all viable
candidates and analyze if they lead to sensible theories. However,
the availability of well-defined Hamiltonian constraint operators
is by itself a notable technical success. For example, the
analogous problem in quantum geometrodynamics ---a satisfactory
regularization of the Wheeler-DeWitt equation--- is still open
although the formal equation was written down some thirty five
years ago. To be specific, I will focus on the procedure developed
by Rovelli, Smolin, Lewandowski and others which culminated in a
specific construction due to Thiemann \cite{qg5}.

Step ii) has been completed for the Gauss and the co-vector
constraints \cite{qg4}. The mathematical implementation required a
very substantial extension \cite{qg4,books} of the algebraic
quantization program initiated by Dirac, and the use of the
spin-network machinery \cite{qg2} of quantum geometry. Again, the
detailed implementation is a non-trivial technical success and the
analogous task has not been completed in geometrodynamics because
of difficulties associated with infinite dimensional spaces.
Thiemann's quantum scalar constraint is defined \textit{on the
space of solutions to the Gauss and co-vector constraints}. The
problem of finding a general solution to the scalar constraint has
been systematically reduced to that of finding \textit{elementary}
solutions, a task that requires only analysis of linear operators
on certain finite dimensional spaces. In this sense, step ii) has
been completed for all constraints.

This is a striking result. However, it is still unclear whether
this theory is physically satisfactory; at this stage, it is in
principle possible that it captures only an `exotic' sector of
quantum gravity. \textit{A key open problem in loop quantum
gravity is to show that the scalar/Hamiltonian constraint}
---either Thiemann's or an alternative such as the one of Gambini
and Pullin \cite{qg5}--- admits a `sufficient number' of
semi-classical states. Progress on this problem has been slow
because, as explained in Section \ref{s1}, the general issue of
semi-classical limits is itself difficult in background
independent approaches. However, as discussed in Section
\ref{s4.2} below, a systematic understanding has now begun to
emerge and is providing the `infra-structure' needed to analyze
the key problem mentioned above. More generally, while there are
promising ideas to complete step iii), substantial further work is
necessary to fully solve this problem. Recent advance in quantum
cosmology, described in Section \ref{s3.1}, is an example of
progress in this direction and it provides a strong support for
the Thiemann scheme, but of course only within the limited context
of mini-superspaces.

To summarize, the crux of dynamics in the Hamiltonian approach
lies in quantum constraints. While the quantum Gauss and
co-vector/diffeomorphism constraints have been solved, it is not
clear if any of the proposed strategies to solve the
scalar/Hamiltonian constraint incorporates the familiar low energy
physics.

\textit{Remark:} There has been another concern about this class
of regularizations of the scalar constraint which, however, is
less specific. It stems from the structure of the constraint
algebra. To analyze this issue, the domain of definition of the
scalar constraint had to be extended to \textit{certain} states
which are not diffeomorphism invariant, so that the commutators
could be meaningfully calculated. It was then found that the
commutator between any two Hamiltonian constraints vanishes
identically \cite{qg5}, while in the classical theory, the
corresponding Poisson brackets vanishes only on solutions to the
diffeomorphism constraint. However, it was also shown that the
operator representing the right side of the classical Poisson
bracket \textit{also vanishes} on all the quantum states
considered, including the ones which are not diffeomorphism
invariant. Therefore, while the vanishing of the commutator of the
Hamiltonian constraint was unexpected, this analysis does not
reveal a clear-cut problem with these regularizations.

Furthermore, one can follow this scheme step by step in 2+1
gravity where one knows what the result should be. One can obtain
the `elementary solutions' mentioned above and show that all the
standard quantum states ---including the semi-classical ones---
can be recovered as linear combinations of these elementary ones.
As is almost always the case with constrained systems, there are
\textit{many more solutions} and the `spurious ones' have to be
eliminated by the requirement that the physical norm be finite. In
2+1 gravity, the connection formulation used here naturally leads
to a complete set of Dirac observables and the inner-product can
be essentially fixed by the requirement that they be self-adjoint.
In 3+1 gravity, by contrast, we do not have this luxury and the
problem of constructing the physical inner-product is therefore
much more difficult. However, the concern here is that of weeding
out unwanted solutions rather than having a `sufficient number' of
semi-classical ones, a significantly less serious issue at the
present stage.

\section{Applications of quantum geometry}
\label{s3}

In this section, I will summarize two recent developments that
answer several of the questions raised under first two bullets in
the Introduction.

\subsection{Big bang} \label{s3.1}

Over the last three years, quantum geometry has led to some
striking results of direct physical interest. The first of these
concerns the fate of the big-bang singularity.

Traditionally, in quantum cosmology one has proceeded by first
imposing spatial symmetries ---such as homogeneity and isotropy---
to freeze out all but a finite number of degrees of freedom
\textit{already at the classical level} and then quantizing the
reduced system. In the simplest case, the basic variables of the
reduced classical system are the scale factor $a$ and matter
fields $\phi$. The symmetries imply that space-time curvature goes
as $\sim 1/a^2$ and Einstein's equations predict a big-bang, where
the scale factor goes to zero and the curvature blows up. As
indicated in Section \ref{s1}, this is reminiscent of what happens
to ferro-magnets at the Curie temperature: magnetization goes to
zero and the susceptibility diverges. By analogy, the key question
is: Do these `pathologies' disappear if we re-examine the
situation in the context of an appropriate quantum theory? In
traditional quantum cosmologies, without an additional input, they
do not. That is, typically, to resolve the singularity one either
has to introduce matter with unphysical properties or introduce
boundary conditions, e.g., by invoking new principles.

In a series of seminal papers \cite{bb}, Bojowald has shown that
the situation in loop quantum cosmology is quite different: the
underlying quantum geometry makes a \textit{qualitative}
difference very near the big-bang. In the standard procedure
summarized above, the reduction is carried out at the classical
level and this removes all traces of the fundamental discreteness.
Therefore, the key idea in Bojowald's analysis is to retain the
essential features of quantum geometry by first quantizing the
kinematics of the \textit{full theory} as in Section \ref{s2.2}
and then restricting oneself to \textit{quantum} states which are
spatially homogeneous and isotropic. As a result, the scale factor
operator $\hat{a}$ has \textit{discrete eigenvalues}. The
continuum limit is reached rapidly. For example, the gap between
an eigenvalue of $\hat{a}$ of $\sim 1 {\rm cm}$ and the next one
is less than $\sim 10^{-30}\, \lp$! Nonetheless, near $a \sim \lp$
there are surprises and predictions of loop quantum cosmology are
very different from those of traditional quantum cosmology.

The first surprise occurs already at the kinematical level. Recall
that, in the classical theory curvature is essentially given by
$1/a^2$, and blows up at the big-bang. What is the situation in
quantum theory? Denote the Hilbert space of spatially homogeneous,
isotropic kinematical quantum states by $\Hih$. A self-adjoint
operator $\widehat{\rm curv}$ corresponding to curvature can be
constructed on $\Hih$ and turns out to be \textit{bounded from
above}. This is very surprising because $\Hih$ admits an
eigenstate of the scale factor operator $\hat{a}$ with a discrete,
zero eigenvalue! At first, it may appear that this could happen
only by an artificial trick in the construction of $\widehat{\rm
curv}$ and that this quantization can not possibly be right
because it seems to represent a huge departure from the classical
relation $({\rm curv})\, a^2 =1$. However, these concerns turn out
to be misplaced. The procedure for constructing $\widehat{\rm
curv}$ is natural and, furthermore, descends from the full theory;
Bojowald essentially repeats a key step in Thiemann's procedure of
defining the quantum scalar/Hamiltonian constraint in the full
theory. Let us examine the properties of $\widehat{\rm curv}$. Its
upper bound $u_{\rm curv}$ is finite but absolutely huge:
\be \label{3.1} u_{\rm curv} \, \sim \,\frac{256}{81}\,
\frac{1}{\lp^2} \equiv \frac{256}{81}\, \frac{1}{G\hbar} \ee
or, about $10^{77}$ times the curvature at the horizon of a solar mass
black hole. The functional form of the upper bound is also
illuminating. Recall that, in the case of an hydrogen atom, energy is
unbounded from below classically but, thanks to $\hbar$, we obtain a
finite value, $E_0 = -\, (me^4/\hbar^2)$, in quantum
theory. Similarly, $u_{\rm curv}$ is finite because $\hbar$ is
non-zero and tends to the classical answer as $\hbar$ tends to
zero. At curvatures as large as $u_{\rm curv}$, it is natural to
expect large departures from classical relations such as $({\rm
curv})\, a^2 =1$. But is this relation recovered in the semi-classical
regime? The answer is in the affirmative. In fact it is somewhat
surprising how quickly this happens. As one would expect, one can
simultaneously diagonalize $\hat{a}$ and $\sqrt{\widehat{\rm curv}}$. If we
denote their eigenvalues by $a_n$ and $b_n$ respectively, then
$(a_n\cdot b_n -1)$ is of the order $10^{-4}$ at $n =100$ and decreases
rapidly as $n$ increases.  These properties show that, in spite of the
initial surprise, the quantization procedure is viable. Furthermore,
one can apply it also to more familiar systems such as a particle
moving on a circle and obtain results which at first seem surprising
but are in complete agreement with the standard quantum theory of
these systems.

\begin{figure}[ht]
\begin{center}
 \includegraphics[width=12cm,height=8cm,keepaspectratio]{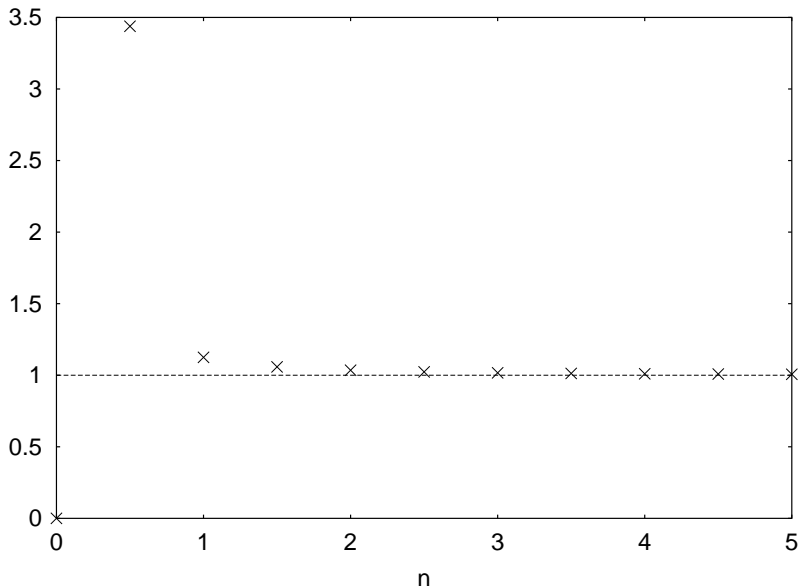}
\end{center}
\caption{The product $a_n \cdot b_n $ as a function of $n$. The dashed 
line is the classical value of $a\cdot \sqrt{\rm curv}$.} 
\label{prod}
\end{figure}

Since the curvature is bounded above in the entire Hilbert space,
one might hope that the quantum evolution may be well-defined
right through the big-bang singularity. Is this in fact the case?
The second surprise is that although the quantum evolution is
close to that of the Wheeler-DeWitt equation of standard quantum
cosmology for large $a$, there are dramatic differences near the
big-bang which makes it well defined even \textit{at} the
big-bang, without any additional input. As one might expect, the
`evolution' is dictated by the quantum scalar constraint operator.
To obtain this operator, Bojowald again follows, step by step, the
procedure used by Thiemann in the full theory. Let us expand out
the full quantum state as $\mid\Psi\!> = \sum_n \psi_n (\phi) \mid
n\!>$ where $\mid n\!>$ are the eigenstates of the scale factor
operator and $\phi$ denotes matter fields. Then, the scalar
constraint takes the form:
\be \label{3.2} c_n \psi_{n+8}(\phi) + d_{n} \psi_{n+4} (\phi) +
e_n \psi_{n} (\phi)+ f_n \psi_{n-4} (\phi) + g_n \psi_{n-8}(\phi)
\, = \, \gamma \lp^2 \,\,\hat{H}_\phi \psi_n(\phi) \ee
where $c_n,\ldots g_n$ are fixed numerical coefficients, $\gamma$
the Barbero-Immirzi parameter and $\hat{H}_\phi$ is the matter
Hamiltonian. (Again, using the Thiemann regularization, one can
show that the matter Hamiltonian is a well-defined operator.)
Primarily, being a constraint equation, (\ref{3.2}) constrains the
physically permissible $\psi_n(\phi)$. However, \textit{if} we
choose to interpret the scale factor (more precisely, the square
of the scale factor times the determinant of the triad) as a time
variable, (\ref{3.2}) can be interpreted as an `evolution
equation' which evolves the state through discrete time steps. In
a (large) neighborhood of the big-bang singularity, this
`deparametrization' is viable. For the choice of factor ordering
used in the Thiemann regularization, one can evolve in the past
through $n=0$, i.e. right through the classical singularity. Thus,
the infinities predicted by the classical theory at the big-bang
are artifacts of assuming that the classical, continuum space-time
approximation is valid right up to the big-bang. In the quantum
theory, the state can be evolved through the big-bang without any
difficulty. However, the classical space-time description fails
near the big-bang; quantum evolution is well-defined but the
classical space-time `dissolves'.

The `evolution' equation (\ref{3.2}) has other interesting
features. To begin with, the space of solutions is 16 dimensional.
Can we single out a preferred solution by imposing a
\textit{physical} condition? One possibility is to impose
\textit{pre-classicality}, i.e., to require that the quantum state
not oscillate rapidly from one step to the next at \textit{late}
times when we know our universe behaves classically. Although this
is an extra input, it is not a theoretical prejudice about what
should happen at (or near) the big-bang but an observationally
motivated condition that is clearly satisfied by our universe. The
coefficients $c_n,\ldots g_n$ of (\ref{3.2}) are such that this
condition singles out a solution uniquely. One can ask what this
state does at negative times, i.e., before the big-bang. (Time
becomes negative because triads flip orientation on the `other
side'.) Preliminary indications are that the state does not become
pre-classical there. If this is borne out by detailed
calculations, then the qualitative analogy with ferro-magnets
would become closer, our side of the big-bang being analogous to
the ferro-magnetic phase in which classical geometry (the analog
of magnetization) is both meaningful and useful and the `other'
side being analogous to the para-magnetic phase where it is not.
Another interesting feature is that the standard Wheeler-DeWitt
equation is recovered if we take the limit $\gamma \to 0$ and $n
\to \infty$ such that the eigenvalues of $\hat{a}$ take on
continuous values. This is completely parallel to the limit we
often take to coarse grain the quantum description of a rotor to
`wash out' discreteness in angular momentum eigenvalues and arrive
at the classical description. From this perspective, then, one is
led to say that the most striking of the consequences of loop
quantum gravity are not seen in standard quantum cosmology because
it `washes out' the fundamental discreteness of quantum geometry.

Finally, the detailed calculations have revealed another
surprising feature. The fact that the quantum effects become
prominent near the big bang, completely invalidating the classical
predictions, is pleasing but not unexpected. However, prior to
these calculations, it was not clear how soon after the big-bang
one can start trusting semi-classical notions and calculations. It
would not have been surprising if we had to wait till the radius
of the universe became, say, a few million times the Planck
length. These calculations strongly suggest that a few tens of
Planck lengths should suffice. This is fortunate because it is now
feasible to develop quantum numerical relativity; with
computational resources commonly available, grids with $(10^6)^3$
points are hopelessly large but one with $(100)^3$ points could be
manageable.

\subsection{Black-holes}
\label{s3.2}

Loop quantum cosmology illuminates dynamical ramifications of
quantum geometry but within the context of mini-superspaces where
all but a finite number of degrees of freedom are frozen. In this
sub-section, I will discuss a complementary application where one
considers the full theory but probes consequences of quantum
geometry which are not sensitive to full quantum dynamics ---the
application of the framework to the problem of black hole entropy.
This discussion is based on joint work with Baez, Corichi and
Krasnov \cite{bh} which itself was motivated by earlier work of
Krasnov, Rovelli and others.

As explained in the Introduction, since mid-seventies, a key
question in the subject has been: What is the statistical
mechanical origin of the black hole entropy $S_{\rm BH} = ({a_{\rm
hor}/ 4\lp^2})$? What are the microscopic degrees of freedom that
account for this entropy? This relation implies that a solar mass
black hole must have $(\exp 10^{77})$ quantum states, a number
that is \textit{huge} even by the standards of statistical
mechanics. Where do all these states reside? To answer these
questions, in the early nineties Wheeler had suggested the
following heuristic picture, which he christened `It from Bit'.
Divide the black hole horizon in to elementary cells, each with
one Planck unit of area, $\lp^2$ and assign to each cell two
microstates, or one `bit'. Then the total number of states ${\cal
N}$ is given by ${\cal N} = 2^n$ where $n = ({a_{\rm hor}/
\lp^2})$ is the number of elementary cells, whence entropy is
given by $S = \ln {\cal N} \sim a_{\rm hor}$. Thus, apart from a
numerical coefficient, the entropy (`It') is accounted for by
assigning two states (`Bit') to each elementary cell. This
qualitative picture is simple and attractive. But the key open
issue was: can these heuristic ideas be supported by a systematic
analysis from first principles? Quantum geometry has supplied the
required analysis.%
\footnote{However, I should add that this account does not follow
chronology. Black hole entropy was computed in quantum geometry
quite independently and the fact that the `It from Bit' picture
works so well in the final picture came as a surprise.}

A systematic approach requires that we first specify the class of
black holes of interest. Since the entropy formula is expected to
hold unambiguously for black holes in equilibrium, most analyses
were confined to \textit{stationary}, eternal black holes (i.e.,
in 4-dimensional general relativity, to the Kerr-Newman family).
From a physical viewpoint however, this assumption seems overly
restrictive. After all, in statistical mechanical calculations of
entropy of ordinary systems, one only has to assume that the given
system is in equilibrium, not the whole world. Therefore, it
should suffice for us to assume that the black hole itself is in
equilibrium; the exterior geometry should not be forced to be
time-independent. Furthermore, the analysis should also account
for entropy of black holes which may be distorted or carry
(Yang-Mills and other) hair. Finally, it has been known since the
mid-seventies that the thermodynamical considerations apply not
only to black holes but also to cosmological horizons. A natural
question is: Can these diverse situations be treated in a single
stroke?  Within the quantum geometry approach, the answer is in
the affirmative. The entropy calculations have been carried out in
the recently developed framework of `isolated horizons'  which
encompasses all these situations. Isolated horizons serve as
`internal boundaries' whose intrinsic geometries (and matter
fields) are time-independent, although space-time geometry as well
as matter fields in the external space-time region can be fully
dynamical. The zeroth and first laws of black hole mechanics have
been extended to isolated horizons. Entropy associated with an
isolated horizon refers to the family of observers in the exterior
for whom the isolated horizon is a physical boundary that
separates the region which is accessible to them from the one
which is not. This point is especially important for cosmological
horizons where, without reference to observers, one can not even
define horizons. States which contribute to this entropy are the
ones which can interact with the states in the exterior; in this
sense, they `reside' on the horizon.

\begin{figure}
\centerline{\hbox{\epsfig{figure=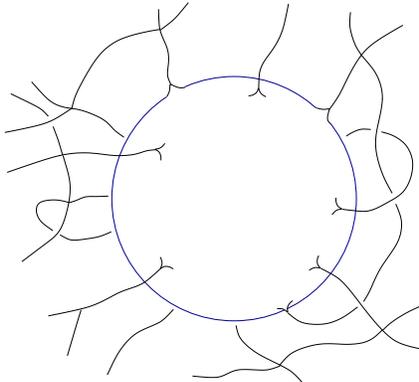,height=2in}}}
\caption{Quantum Horizon. Polymer excitations in the bulk puncture the
horizon, endowing it with quantized area. Intrinsically, the horizon
is flat except at punctures where it acquires a quantized deficit
angle. These angles add up to endow the horizon with a 2-sphere
topology.}
\label{fig2}
\end{figure}

In the detailed analysis, one considers space-times admitting an
isolated horizon as inner boundary and carries out a systematic
quantization. The quantum geometry framework can be naturally extended
to this case. The isolated horizon boundary conditions imply that the
intrinsic geometry of the quantum horizon is described by the so
called $\U(1)$ Chern-Simons theory on the horizon. This is a
well-developed, topological field theory. A deeply satisfying feature
of the analysis is that there is a seamless matching of three
otherwise independent structures: the isolated horizon boundary
conditions, the quantum geometry in the bulk, and the Chern-Simons
theory on the horizon. In particular, one can calculate eigenvalues of
certain physically interesting operators using purely bulk quantum
geometry without any knowledge of the Chern-Simons theory, or using
the Chern-Simons theory without any knowledge of the bulk quantum
geometry. The two theories have never heard of each other. Yet, thanks
to the isolated horizon boundary conditions, the two infinite sets of
numbers match exactly, providing a coherent description of the quantum
horizon.

In this description, the polymer excitations of the bulk geometry,
each labelled by a spin $j_I$, pierce the horizon, endowing it an
elementary area $a_{j_I}$ given by (\ref{2.1}). The sum
$\textstyle{\sum_I} a_{j_I}$ adds up to the total horizon area
$a_{\rm hor}$. The intrinsic geometry of the horizon is flat
except at these puncture, but at each puncture there is a
\textit{quantized} deficit angle. These add up to endow the
horizon with a 2-sphere topology. For a solar mass black hole, a
typical horizon state would have $10^{77}$ punctures, each
contributing a tiny deficit angle. So, although the quantum
geometry \textit{is} distributional, it can be well approximated
by a smooth metric.

The counting of states can be carried out as follows. First one
constructs a micro-canonical ensemble by restricting oneself only
to those states for which the total area, angular momentum, and
charges lie in small intervals around fixed values $a_{\rm hor},
J_{\rm hor}, Q^i_{\rm hor}$. (As is usual in statistical
mechanics, the leading contribution to the entropy is independent
of the precise choice of these small intervals.) For each set of
punctures, one can compute the dimension of the surface Hilbert
space, consisting of Chern-Simons states compatible with that set.
One allows all possible sets of punctures (by varying both the
spin labels and the number of punctures), subject to the
constraint that the total area $a_{\rm hor}$ be fixed, and adds up
the dimensions of the corresponding surface Hilbert spaces to
obtain the number ${\cal N}$ of permissible surface states. One
finds that the horizon entropy $S_{\rm hor}$ is given by
\be \label{3.3} S_{\rm hor} := \ln {\cal N} =
\frac{\gamma_o}{\gamma}\, \frac{a_{\rm hor}}{\lp^2} + {\cal O}
(\frac{\lp^2}{a_{\rm hor}}) ,\quad {\rm where} \quad \gamma_o =
\frac{\ln 2}{\sqrt{3} \pi } \ee
Thus, for large black holes, entropy is indeed proportional to the
horizon area. This is a non-trivial result; for examples, early
calculations often led to proportionality to the square-root of
the area. However, even for large black holes, one obtains
agreement with the Hawking-Bekenstein formula only in the sector
of quantum geometry in which the Barbero-Immirzi parameter
$\gamma$ takes the value $\gamma = \gamma_o$. Thus, while all
$\gamma$ sectors are equivalent classically, the standard quantum
field theory in curved space-times is recovered in the
semi-classical theory only in the $\gamma_o$ sector of quantum
geometry. It is quite remarkable that thermodynamic considerations
involving \textit{large} black holes can be used to fix the
quantization ambiguity which dictates such Planck scale properties
as eigenvalues of geometric operators. Note however that the value
of $\gamma$ can be fixed by demanding agreement with the
semi-classical result just in one case ---e.g., a spherical
horizon with zero charge, or a cosmological horizon in the de
Sitter space-time, or, \ldots. Once the value of $\gamma$ is
fixed, the theory is completely fixed and we can ask: Does this
theory yield the Hawking-Bekenstein value of entropy of
\textit{all} isolated horizons, irrespective of the values of
charges, angular momentum, and cosmological constant, the amount
of distortion, or hair. The answer is in the affirmative. Thus,
the agreement with quantum field theory in curved space-times
holds in \textit{all} these diverse cases.

Why does $\gamma_o$ not depend on other quantities such as
charges? This important property can be traced back to a key
consequence of the isolated horizon boundary conditions: detailed
calculations show that only the gravitational part of the
symplectic structure has a surface term at the horizon; the matter
symplectic structures have only volume terms. (Furthermore, the
gravitational surface term is insensitive to the value of the
cosmological constant.) Consequently, there are no independent
surface quantum states associated with matter. This provides a
natural explanation of the fact that the Hawking-Bekenstein
entropy depends only on the horizon geometry and is independent of
electro-magnetic (or other) charges.

Finally, let us return to Wheeler's `It from Bit'. One can ask:
what are the states that dominate the counting? Perhaps not
surprisingly, they turn out to be the ones which assign to each
puncture the smallest quantum of area (i.e., spin value $j =
\textstyle{1\over 2}$), thereby maximizing the number of
punctures. In these states, each puncture defines Wheeler's
`elementary cell' and his two states correspond to whether the
deficit angle is positive or negative.

To summarize, quantum geometry naturally provides the micro-states
responsible for the huge entropy associated with horizons. In this
analysis, all black holes and cosmological horizons are treated in
an unified fashion; there is no restriction, e.g., to
near-extremal black holes. The sub-leading term has also been
calculated and shown to be proportional to $\ln a_{\rm hor}$.
Finally, in this analysis quantum Einstein's equations
\textit{are} used. In particular, had we not imposed the quantum
Gauss and co-vector/diffeomorphism constraints on surface states,
the spurious gauge degrees of freedom would have given an infinite
entropy. However, because of the isolated horizon boundary
conditions, the scalar/Hamiltonian constraint has to be imposed
just in the bulk. Since in the entropy calculation one traces over
bulk states, the final result is insensitive to the details of how
this (or any other bulk) equation is imposed. Thus, as in other
approaches to black hole entropy, the calculation does not require
a complete knowledge of quantum dynamics.

\section{Current work}
\label{s4}

Work is now in progress along many directions, ranging from the
fate of the `final' black hole singularity in quantum geometry and
the associated issue of `information loss', to predictions of
quantum cosmology for structure formation, to practical methods of
constructing Dirac observables, to possible experimental tests of
quantum geometry. To illustrate these developments, in this
section I will discuss two main thrusts.

\subsection{Spin foams}
\label{s4.1}

Spin foams can be thought of as histories traced out by `time
evolution' of spin networks and provide a path integral approach
to quantum dynamics. I will focus on the fascinating `finiteness
results' \cite{sf} obtained by Crane, Rovelli and especially Perez
based on earlier work also by Baez, Barrett, De Pietri, Freidel,
Krasnov, Mikovi\v{c}, Reisenberger  and others.

In the gravitational context, the path integral can play two
roles. First, as in standard quantum field theories, it can be
used to compute `transitions amplitudes'. However outside, say,
perturbation theory about a background space-time, there still
remain unresolved conceptual questions about the physical meaning
of such amplitudes. The second role is `cleaner': as in the
Euclidean approach of Hawking and others, it can be considered as
a device to extract physical states, i.e. solutions to all the
quantum constraint equations. In this role as an
\textit{extractor}, it can shed new light on the difficult issue
of regularization of the Hamiltonian constraint discussed in
Section \ref{s2.3}.

The well-defined quantum kinematics of Section \ref{s2.2} has
motivated specific proposals for the definition of path integrals,
often called `state sum models'. Perhaps the most successful of
these is the Barrett-Crane model (modified slightly to handle a
technical issue). At the classical level, one regards general
relativity as a topological field theory, \textit{called the} BF
\textit{theory}, supplemented with an algebraic constraint. The BF
theory is itself a generalization of the Chern-Simons theory
mentioned in Section \ref{s3.2} and has been investigated in
detail in the mathematical physics literature. However, the role
of the additional constraint is very important. Indeed, the BF
theory has no local degrees of freedom; it is the extra constraint
that reduces the huge gauge freedom of the BF theory, thereby
recovering the local degrees of freedom of general relativity. The
crux of the problem in quantum gravity is the appropriate
incorporation of this constraint. At the classical level, the
constrained B-F theory is equivalent to general relativity. To
obtain Euclidean general relativity, one has to start with the BF
theory associated with $SO(4)$ while the Lorentzian theory results
if one uses $SO(3,1)$ instead. The Barrett-Crane model is a
specific proposal to define path integrals for the constrained BF
theory in either case.

Fix a 4-manifold $M$ bounded by two 3-manifolds $\S_1$ and $\S_2$.
Spin-network states on the two boundaries can be regarded as
`initial' and `final' quantum geometries. One can then consider
histories, i.e., quantum 4-geometries, joining them. Each history
is a spin-foam. Each vertex of the initial spin-network on $\S_1$
`evolves' to give a 1-dimensional edge in the spin-foam and each
edge, to give a 2-dimensional face. Consequently, each face
carries a spin label $j$. However, in the course of `evolution'
\textit{new vertices} can appear, making the dynamics non-trivial
and yielding a non-trivial amplitude for an `initial' spin-network
with $n_1$ vertices to evolve in to a `final' spin-network with
$n_2$ vertices. For mathematical clarity as well as physical
intuition, it is convenient to group spin-foams associated with
the same 4-dimensional graph but differing from one another in the
labels, such as the spins $j$ carried by faces. Each group is said
to provide a \textit{discretization} of $M$. Physically, a
discretization has essentially just the topological information.
The geometrical information ---such as the area associated with
each face--- resides in the labels. This is a key distinction from
lattice gauge theories with a background metric, where a
discretization itself determines, e.g., the edge lengths and hence
how refined the lattice is.

A key recent development is the discovery that the
non-perturbative path integral, defined by the (modified)
Barrett-Crane model is, in a certain precise sense, equivalent to
a manageable \textit{group field theory} (GFT) in the sense
specified below. The GFT is a rather simple quantum field theory,
defined on four copies of the underlying group  ---$\SL(2,C)$ in
the case of Lorentzian gravity and ${\rm Spin}(4)$ in the case of
Euclidean. (Note that these are just double covers of the Lorentz
group and the rotation group in Euclidean 4-space.) Thus GFTs live
in high dimensions. The action has a `free part' and an
interaction term with a coupling constant $\lambda$. But the free
part is non-standard and does not have the familiar kinetic term,
whence the usual non-renormalizability arguments for higher
dimensional, interacting theories do not apply. In fact, the first
key recent result is that \textit{this GFT is finite order by
order in the Feynman perturbation expansion.} The second key
result is $A_{\rm BC}{(n)}= A_{\rm GFT} {(n)}$, where $A_{\rm
BC}{(n)}$ is the Barret-crane amplitude obtained by summing over
all geometries (i.e., spin labels $j$) for a fixed discretization
and $A_{\rm GFT}{(n)}$ is the coefficient of $\lambda^n$ in the
Feynman expansion of the GFT. Together, the two results imply
that, in this approach to quantum gravity, \textit{sum over
geometries for a fixed discrete topology is finite}. This is a
highly non-trivial result because, on each face, the sum over $j$s
ranges from zero to infinity; there is no cut-off.

These results do not show that the full path integral is finite
because the discrete topology is kept fixed in the sum. Work is in
progress on removing this dependence. But even as they stand, the
results have a striking, qualitative similarity with the
finiteness results that have been recently obtained in other
approaches. The order by order finiteness is reminiscent of the
order by order ultraviolet finiteness of string perturbation
theory. In the present case, the situation is somewhat better:
there is both ultra-violet \textit{and} infra-red finiteness.
Qualitatively, the first is ensured by the discreteness of
underlying geometry while the second by the fact that the sum
converges even though arbitrarily large values of $j$s are
allowed. The second result ---equivalence of the GFT with a
certain definition of quantum gravity--- is reminiscent of the
conjectures discussed by Maldacena in this conference. In both
cases, there is a mathematical equivalence between quantum gravity
and certain field theories which have no knowledge of the physical
space-time. On the one hand, results are stronger in the present
case: both sides of the equality are separately defined and the
thrust is on proving the equality, not just on constructing
evidence in support of it. However, in contrast to Maldacena's
bold conjecture, here the equality in question is only order by
order. Finally, recently Luscher and Reuter have found evidence
for nonperturbative renormalizability of 4-dimensional Euclidean
quantum general relativity (stemming from the existence of a
non-trivial fixed point) \cite{sf}. It is natural to ask: Is there
a relation to Perez's finiteness results in the Euclidean
signature?

\subsection{Relation to low energy physics}
\label{s4.2}

The basic mathematical structures underlying loop quantum gravity
are very different from those used in the text-book treatments of
low energy physics. For example, in quantum geometry the
fundamental excitations are one dimensional, polymer-like; a
convenient basis of states is provided by spin-networks; and,
eigenvalues of the basic geometric operators are discrete.  By
contrast, in the Fock framework of low energy physics the
fundamental excitations are 3-dimensional, wavy; the convenient
basis is labelled by the number of particles, their momenta and
helicities; and, all geometric operators have continuous spectra.
The challenge is to bridge the gap between these apparently
disparate frameworks. These differences are inevitable, given that
the standard quantum field theory is constructed in Minkowski
space, while loop quantum gravity does not refer to any background
geometry. Nonetheless, if loop quantum gravity is correct, the
standard Fock framework should emerge in a suitable semi-classical
approximation. Over the last decade, semi-classical states
approximating classical geometries have been constructed with
various degrees of precision \cite{sc}. However, the relation to
Fock quantization has been explored only recently \cite{fock}. My
summary of this development will be even briefer than that of
other advances because this topic was discussed in some detail in
Rovelli's workshop D1i.

The recent developments are based on a key mathematical insight
due to Varadarajan. Fock states do not belong to the Hilbert space
$\H$ of polymer excitations. However, neither do the physical
states, i.e., the solutions to quantum constraints. This is not a
peculiarity of general relativity. Even for quantum mechanical
systems such as a free particle in Minkowski space, the physical
states belong not to the kinematical Hilbert space, but to a
natural extension of it, constructed from the refinement of the
Dirac quantization program for constrained systems, mentioned in
Section \ref{s2.3}. Varadarajan showed that Fock states do belong
to this extension ${\E}$ of $\H$. Thus, the natural habitat $\E$
for the physical states in loop quantum gravity also accommodates
the standard Fock states of low energy physics.

At first, this result seems very surprising. For example, in the
polymer description of a quantum Maxwell field, fluxes of electric
field are quantized. On the Fock space, on the other hand, these
flux operators are not even defined unless surfaces are thickened
and, when they are, the quantization is lost. This phenomenon
succinctly captures the tension between the polymer and Fock
excitations. It turns out that the flux operators are indeed
well-defined on $\E$ but they do not map the Fock space to itself;
flux quantization is lost because the Fock space is `too small'
and fails to accommodate any of the eigenvectors of these
operators with discrete eigenvalues. The situation with quantum
geometry is completely analogous: The discreteness of quantum
geometry is lost if one insists that \textit{all} quantum gravity
states must reside in the Fock space of gravitons. To see the
discreteness, one has to allow states which are `purely quantum
mechanical' and can not be regarded as excitations on \textit{any}
classical background geometry.

The interplay between the Fock  and polymer excitations can now be
studied in detail, thanks to a key property: each Fock state casts
a `shadow' in each finite dimensional space $\H_{\gamma, \vec{j}}$
of $\H$ and can be fully recovered from the collection of these
shadows. Using these shadows one can analyze a number of features
of the Fock framework, such as construction of coherent states,
expectations values and fluctuations of fields, needed in the
analysis of semi-classical issues. Thus, one can now begin to
analyze two key questions: Can the background-independent,
non-perturbative theory reproduce the familiar low-energy physics
on suitable coarse graining?  and, Can one pin-point where and why
the standard perturbation theory fails?

\section{Conclusion}
\label{s5}

In the last two sections, I have summarized recent advances which
have answered some of the long standing questions of quantum
gravity raised in the Introduction. Personally, I find it very
satisfying that a number of the key ideas came from younger
researchers  ---from Bojowald in quantum cosmology, from Krasnov
in the understanding of quantum black holes, Perez in spin foams
and Varadarajan in the relation to low energy physics.

Throughout the development of loop quantum gravity, unforeseen
simplifications have arisen regularly, leading to surprising
solutions to seemingly impossible difficulties. Progress could
occur because some of the obstinate problems which had slowed
developments in background independent approaches, sometimes for
decades, evaporated when `right' perspectives were found. I will
conclude with a few examples.

$\bullet$ Up until the early nineties, it was widely believed that
spaces of connections do not admit non-trivial diffeomorphism
invariant measures. This would have made it impossible to develop
a background independent approach. Quite surprisingly, such a
measure could be found by looking at connections from a slightly
more general perspective. It is simple, natural, and has just the
right structure to support quantum geometry. This geometry, in
turn, supplied some missing links, e.g., by providing just the
right expressions that Ponzano-Regge had to postulate without
justification in their celebrated, early work.

$\bullet$ Fundamental discreteness first appeared in a startling
fashion in the construction of the so-called weave states, which
approximate a classical 3-geometry. In this construction, the
polymer excitations were introduced as a starting point with the
goal of taking the standard continuum limit. It came as a major
surprise that, if one wants to recover a given classical geometry
\textit{on large scales}, one can not take this limit, i.e., one
can not pack the polymer excitations arbitrarily close together;
there is an in-built discreteness.

$\bullet$ At a heuristic level, it was found that the Wilson loop
functionals of a suitably defined connection around a smooth loop
solve the notoriously difficult quantum scalar constraint
automatically. No one had anticipated, even heuristically, such
simple and natural solutions. This calculation suggested that the
action of the constraint operator is concentrated at `nodes',
i.e., intersections, which in turn led to strategies for its
regularization.

$\bullet$ As I indicated in some detail, unforeseen insights arose
in the well-studied subject of quantum cosmology essentially by
taking an adequate account of the quantum nature of geometry,
i.e., by respecting the fundamental discreteness of the
eigenvalues of the scale factor operator. Similarly, in the case
of black holes, three quite distinct structures ---the isolated
horizon boundary conditions, the bulk quantum geometry and the
surface Chern-Simons theory--- blended together unexpectedly to
provide a coherent theory of quantum horizons.

Repeated occurrence of such `unreasonable' simplifications
suggests that the ideas underlying loop quantum gravity may have
captured an essential germ of truth.

\bigskip

\noindent\textbf{Acknowledgements} I would like to thank Martin
Bjowald, Jerzy Lewandowski and Alejandro Perez for their comments
on the manuscript and Carlo Rovelli for correspondence. This work
was supported in part by the NSF grant PHY-0090091 and the Eberly
research funds of Penn State.


\begin{thebibliography}{99}


\bibitem{books} \textsl{Books:}\\
A.\ Ashtekar, {\it Lectures on non-perturbative canonical
gravity,} (World Scientific, Singapore, 1991)\\
R.\ Gambini and J.\ Pullin, {\em Loops, knots, gauge theories and
quantum gravity}, (CUP, Cambridge, 1996)

\bibitem{class} \textsl{Classical theory:}\\
A.\ Ashtekar, New variables for classical and quantum gravity,
Phys.\ Rev.\ Lett.\ {\bf 57}, 2244-2247, (1986)\\
New Hamiltonian formulation of general relativity,
Phys.\ Rev.\ {\bf D36}, 1587-1602 (1987) \\
A.\ Ashtekar, J.\ D.\ Romano and R.\ S.\ Tate, New variables for
gravity: Inclusion of matter, Phys. rev. {\bf D40}, 2572-2587
(1989)\\
F.\ Barbero, Real Ashtekar variables for Lorentzian signature
space-times, Phys.\ Rev.\ {\bf D51}, 5507-5510 (1996)

\bibitem{loop} \textsl{Connctions and loops:}\\
C.\ Rovelli and L. Smolin, Loop representation for quantum
general relativity, Nucl.\ Phys.\ {\bf B331} 80-152 (1990)\\
A.\ Ashtekar, V.\ Husain, C.\ Rovelli, J.\ Samuel and L.\ Smolin,
2+1 quantum gravity as a toy model for the 3+1 theory, Class.\
Quant.\ Grav. {\bf 6}, L185-L193 (1989)\\
A.\ Ashtekar, C.\ Rovelli and L.\ Smolin, Gravitons and loops,
Phys.\ Rev.\ {\bf D44}, 1740-1755 (1991).

\bibitem{qg1} \textsl{Quantum Geometry: Basics}\\
A.\ Ashtekar and C.\ J.\ Isham, Representation of the holonomy
algebras of gravity and non-Abelian gauge theories, Class.\
Quant.\ Grav.\ {\bf 9}, 1433-1467 (1992)\\
A.\ Ashtekar and J.\ Lewandowski, Representation theory of
analytic holonomy algebras, in {\sl Knots and Quantum Gravity},
edited by J.\ C.\ Baez, Oxford U.\ Press, Oxford, (1994)\\
J.\ C.\ Baez, Generalized measures in gauge theory, Lett.\ Math.\
Phys.\ {\bf 31}, 213-223 (1994)\\
D.\ Marolf and J.\ Mour\~ao, On the support of the
Ashtekar-Lewandowski measure, Commun.\ Math.\ Phys.\ {\bf 170} ,
583-606 (1995)\\
A.\ Ashtekar and J.\ Lewandowski, Projective techniques and
functional integration, Jour.\ Math.\ Phys.\ {\bf 36}, 2170-2191
(1995)\\
J.\ C.\ Baez and S.\ Sawin, Functional integration on spaces of
connections, Jour.\ Funct.\ Analysis {\bf 150}, 1-27 (1997)\\


\bibitem{qg2} \textsl{spin networks:}\\
R.\ Penrose, Angular momentum: an approach to combinatorial
space-time, in {\sl Quantum Theory and Beyond}, edited by Ted
Bastin, (Cambridge University Press, 1971)\\
C.\ Rovelli and L.\ Smolin, Spin networks and quantum gravity,
Phys.\ Rev.\ {\bf D52}, 5743-5759 (1995)\\
J.\ C.\ Baez, Spin networks in non-perturbative quantum gravity,
in {\sl The Interface of Knots and Physics,} edited by L.\
Kauffman, American Mathematical Society, Providence,
1996, pp.\ 167-203 \\
J.\ C.\ Baez, Spin networks in gauge theory, Adv. Math. {\bf 117},
253-272 (1996)\\
J.\ C.\ Baez and S.\ Sawin, Diffeomorphism-invariant spin network
states, Jour.\ Funct.\ Analysis {\bf 158}, 253-266 (1998)

\bibitem{qg3} \textsl{Geometric operators and their properties}\\
C.\ Rovelli and L.\ Smolin, Discreteness of area and volume in
quantum gravity, Nucl.\ Phys. {\bf B442}, 593-622 (1995); Erratum:
Nucl.\ Phys. {\bf B456}, 753 (1995)\\
R.\ Loll, The volume operator in discretized quantum gravity,
Phys.\ Rev.\ Lett.\ {\bf 75} 3048-3051 (1995) \\
A.\ Ashtekar and J.\ Lewandowski, Differential geometry on the
space of connections using projective techniques, Jour.\
Geo. \& \ Phys. {\bf 17}, 191-230 (1995)\\
A.\ Ashtekar and J.\ Lewandowski, Quantum theory of geometry I:
Area operators, Class.\ Quant.\ Grav.\ {\bf 14}, A55-A81 (1997)\\
Quantum theory of geometry II: Volume Operators, Adv.\ Theo.\
Math.\ Phys.\ {\bf 1}, 388-429
(1997)\\
R.\ Loll, Further results on geometric operators in quantum
gravity, Class.\ Quant.\ Grav. {\bf 14} 1725-1741 (1997)\\
T.\ Thiemann, A length operator for canonical quantum gravity,
Jour.\ Math.\ Phys.\ {\bf 39}, 3372-3392 (1998)\\


\bibitem{bi}\textsl{Barbero-Immirzi ambiguity}\\
G.\ Immirzi, Quantum gravity and Regge calculus, Nucl.\ Phys.\
Proc.\ Suppll. {\bf 57}, 65-72 (1997)\\
C.\ Rovelli and T.\ Thiemann, The Immirizi parameter in quantum
general relativity, Phys.\ Rev.\ {\bf D57} 1009-1014 (1998)\\
R.\ Gambini O.\ Obregon and J.\ Pullin, Yang-Mills analogs of the
Immirzi ambiguity, Phys.\ Rev.\ {\bf D59} 047505 (1999)

\bibitem{qg4} \textsl{Gauss and difeomorphism constraints}\\
D.\ Marolf, Refined algebraic quantization: Systems with a single
constraint,  \texttt{gr-qc/9508015}\\
A.\ Ashtekar, J.\ Lewandowski, D.\ Marolf, J.\ Mour\~ao,  and T.\
Thiemann, Quantization of diffeomorphism invariant theories of
connections with local degrees of freedom, Jour.\ Math.\ Phys.\
{\bf 36}, 6456-6493 (1995)\\
J.\ Lewandowski and T.\ Thiemann, Diffeomorphism invariant quantum
field theories of connections in terms of webs,  Class.\ Quant.\
Grav.\ {\bf 16}, 2299-2322 (1999)\\

\bibitem{qg5}\textsl{Hamiltonian constraint}\\
C.\ Rovelli and L.\ Smolin, The physical Hamiltonian in
nonperturbative quantum gravity, Phys.\ Rev.\ Lett.\ {\bf 72},
446-449 (1994)\\
T.\ Thiemann, Anomaly-free formulation of non-perturbative,
four-dimensional Lorentzian quantum gravity, Phys.\ Lett.\ {\bf
B380} 257-264 (1996)\\
Quantum Spin Dynamics (QSD), Class.\ Quant.\ Grav.\ {\bf 15}
839-873 (1998)\\
QSD III : Quantum Constraint Algebra and Physical Scalar Product
in Quantum General Relativity, Class.\ Quant.\ Grav. {\bf 15}
1207-1247 (1998)\\
QSD V : Quantum Gravity as the Natural Regulator of Matter Quantum
Field Theories, Class.\ Quant.\ Grav. {\bf 15} 1281-1314 (1998)\\
R.\ Gambini, J.\ Lewandowski, D.\ Marolf and J.\ Pullin, On the
consistency of the constraint algebra in spin network quantum
gravity, Int.\ J.\ Mod.\ Phys.\  {\bf D7}, 97-109 (1998)\\
J.\ Lewandowski and D.\ Marolf,  Loop constraints: A habitat and
their algebra, Int.\ J.\ Mod.\ Phys. {\bf D7}, 299-330 (1998)\\
C.\ Di Bartolo, R.\ Gambini, J.\ Griego, J.\ Pullin, Consistent
canonical quantization of general relativity in the space of
Vassiliev knot invariants, Phys.\ Rev.\ Lett. {\bf 84}, 2314-2317
(2000)


\bibitem{bb}\textsl{Big-bang:}\\
M.\ Bojowald, absence of singularity in loop quantum
cosmology, Phys.\ Rev.\ Lett. {\bf 86}, 5227-5230 (2001) \\
Dynamical initial conditions in quantum cosmology,
{\em Phys.\ Rev.\ Lett.} {\bf 87},  121301 (2001)\\
Inverse scale factor in isotropic quantum geometry, { Phys.\ Rev.}
{\bf D64} 084018 (2001)\\
Loop quantum cosmology III: Wheeler-DeWitt operators, {Class.\
Quant.\ Grav.} {\bf 18} 1055--1070 (2001) \\
Loop Quantum Cosmology IV: Discrete Time Evolution, {Class.\
Quant.\ Grav.} {\bf 18} 1071--1088 (2001) \\
{\em Quantum Geometry and Symmetry}, (Shaker-Verlag, Aachen, 2001)

\bibitem{bh} \textsl{Black holes:}\\
A.~Ashtekar, J.~Baez, A.~Corichi, K.~Krasnov, Quantum geometry and
black hole entropy, {Phys.\ Rev.\ Lett.} \textbf{80} 904-907 (1998)\\
A.~Ashtekar, A.~Corichi and K.~Krasnov, Isolated horizons: the
classical phase space, {Adv.\ Theor.\ Math.\
Phys.} \textbf{3}, 418-471 (1999)\\
A.~Ashtekar, J.~Baez, K.~Krasnov, Quantum geometry of isolated
horizons and black hole entropy, {Adv.\ Theo.\ Math.\
Phys.}, \textbf{4}, 1-95 (2000)\\
R.\ K.\ Kaul and P.\ Majumdar, Logarithmic corrections to the
Bekenstein-Hawking entropy, Phys.\ Rev.\ Lett.\ {\bf 84},
5255-5257 (2000)

\bibitem{sf} \textsl{Spin foams and finiteness:}\\
A.~Perez, {Finiteness of a spinfoam model for Euclidean quantum
general  relativity}, Nucl.\ Phys.\ {\bf B599}, 427 (2001)\\
L.~Crane, A.~Perez, and C.~ Rovelli, {Perturbative finiteness in
spin-foam quantum gravity}, Phys.\ Rev.\ Lett.\ {\bf 87} 181301, (2001)\\
A.~Perez and C.~Rovelli, {Spin foam model for Lorentzian general
relativity}, Phys.\ Rev.\  {\bf D63}, 041501 (2001)\\
J.~W.~Barrett, L.~Crane, {Relativistic spin networks and
quantum gravity}, J. Math. Phys. {\bf 39}, 3296 (1998)\\
J.~Baez, {Spin foam models}, Class.\ Quant.\ Grav. {\bf 15},
1827-1858 (1998) \\
J.~W.~ Barrett, L.~Crane, {A Lorentzian signature model for
quantum general relativity}, Class.\ Quant.\ Grav. {\bf 17}
3101-3118 (2000)\\
O.~Lauscher and M.~Reuter, Is quantum Einstein gravity
non-perturbatively renormalizable?, \texttt{hep-th/0110021}

\bibitem{sc} \textsl{Semi-classical states:}\\
A.\ Ashtekar, C.\ Rovelli and L.\ Smolin, Weaving a classical
geometry with quantum threads, Phy.\ Rev.\ Lett.\ {\bf 69},
237-240 (1992)\\
M.\ Arnsdorf, S.\ Gupta, Loop quantum gravity on non-compact
spaces,  Nucl.\ Phys.\ \textbf{ B577} 529-546 (2000)\\
T.\ Thiemann, Gauge field theory coherent states (GCS) : I.
general properties, Class.\ Quant.\ Grav.\ {\bf 18} 2025-2064
(2001)\\
H.\ Sahlmann, T.\ Thiemann and O.\ Winkler, Coherent states for
canonical quantum general relativity and the infinite tensor
product extension, \texttt{gr-qc/0102038}

\bibitem{fock} \textsl{Fock states in the polymer picture:}\\
M.\ Varadarajan, Fock representations from U(1) holonomy
algebras, Phys.\ Rev.\ {\bf D61}, 104001 (2000)\\
M.\ Varadarajan, Photons from quantized electric flux
representations, \texttt{gr-qc/0104051}\\
A.\ Ashtekar and J.\ Lewandowski, Relation between plymer and Fock
excitations, Class.\ Quant.\ Grav.\  {\bf 18}, L117-L127 (2001) \\
L.\ Bombelli, Statistical geometry of random weave states,
\texttt{gr-qc/0101080}.


\end{thebibliography}
\end{document}